# Melting of electronic/excitonic crystals in 2D semiconductor moiré patterns: a perspective from the Lindemann criterion

Jiyong Zhou[1], Jianju Tang[1], Hongyi Yu[1,2*]

[1] Guangdong Provincial Key Laboratory of Quantum Metrology and Sensing & School of Physics and Astronomy, Sun Yat-Sen University (Zhuhai Campus), Zhuhai 519082, China
[2] State Key Laboratory of Optoelectronic Materials and Technologies, Sun Yat-Sen University (Guangzhou Campus), Guangzhou 510275, China
* E-mail: yuhy33@mail.sysu.edu.cn

**Abstract:** Using the Lindemann criterion, we analyzed the quantum and thermal melting of electronic/excitonic crystals recently discovered in 2D semiconductor moiré patterns. We show that the finite 2D screening of the atomically thin material can suppress (enhance) the inter-site Coulomb (dipolar) interaction strength, thus inhibits (facilitates) the formation of the electronic (excitonic) crystal. Meanwhile, a strong enough moiré confinement is found to be essential for realizing the crystal phase with a wavelength near 10 nm or shorter. From the calculated Lindemann ratio which quantifies the fluctuation of the site displacement, we estimate that the crystal will melt into a liquid above a critical temperature ranging from several tens Kelvin to above 100 K (depending on system parameters).

The formation of long-wavelength moiré patterns in van der Waals stacking of 2D semiconducting transition metal dichalcogenides (TMDs) has introduced a new platform for studying exotic quantum phenomena[1,2]. In the past few years, experiments have detected various electronic correlated insulators[3-15] and moiré confined interlayer excitons[16-19] in these systems, which come from the enhanced Coulomb interaction by the 2D geometry combined with the presence of a moiré superlattice potential. By continuously tuning the doping density of the bilayer TMDs moiré system, a variety of quantum electronic crystals with different lattice types have been detected under low temperatures. Besides the triangular Mott insulators at a filling factor of one electron per moiré supercell ($v = 1$), the generalized Wigner crystals and stripe crystals under fractional fillings $v < 1$ as well as the monolayer and bilayer Wigner crystals with the absence of moiré patterns have also been observed[4,8,9,11,13,15], signifying the long-range nature of the Coulomb interaction. Meanwhile in a bilayer system, an electron-hole pair in opposite layers can bound into an interlayer exciton (IX) which has a permanent electric dipole perpendicular to the 2D plane[20]. The dipolar repulsion between IXs trapped in different moiré potential minima can give rise to the formation of a quantum excitonic crystal[21-23], as being observed in recent experiments[24-27].

The phase transition between the crystal and liquid comes from the competition between the kinetic and potential energies, which favor delocalization and localization, respectively. A melting of the crystal will happen when the fluctuation of the crystal site becomes sufficiently large compared to the inter-site distance, which can be realized by the increase of the temperature $T$ (thermal melting) or the decrease of the inter-site distance (quantum melting). This leads to the traditional Lindemann criterion, which states that the melting occurs when the Lindemann ratio $\eta \equiv \sqrt{\langle \mathbf{r}^2 \rangle}/\lambda$ exceeds some critical value $\eta_c$. Here $\langle \mathbf{r}^2 \rangle$ is the mean-square displacement of the crystal site from both the quantum and thermal fluctuations, and $\lambda$ is the distance between the nearest-neighbor sites. The traditional Lindemann criterion is rather successful in describing the melting of 3D atomic crystals (with $\eta_c \approx 0.1$). Meanwhile numerical analyses have shown that it also applies to the quantum melting of 2D quantum crystals under $T = 0$ (with $\eta_c$ between 0.2 to 0.25) [28-30]. However, it fails to describe the thermal melting of 2D crystals, as $\langle \mathbf{r}^2 \rangle$ diverges logarithmically with the area of the 2D system when $T > 0$. Alternatively, one can consider the fluctuation of the relative displacement $\langle (\mathbf{r}_n - \mathbf{r}_{n'})^2 \rangle$ between a pair of nearest-neighbor sites $(n, n')$, and define a modified Lindemann ratio $\eta^{(m)} \equiv \sqrt{\langle (\mathbf{r}_n - \mathbf{r}_{n'})^2 \rangle}/\lambda$ which is finite under $T > 0$ [31]. The modified Lindemann criterion states that the crystal melting occurs when $\eta^{(m)}$ exceeds some critical value $\eta_c^{(m)} \approx 0.31$ for 2D quantum crystals formed by the Coulomb or dipolar repulsions[32]. It should be emphasized that although the Lindemann criterion has been confirmed by numerical simulations and experiments, the obtained empirical constants $\eta_c$ and $\eta_c^{(m)}$ from different literatures have slight variations. Nevertheless, the melting process can be qualitatively analyzed from the dependences of $\eta$ and $\eta^{(m)}$ with system parameters. In this work, we calculate $\eta$ and $\eta^{(m)}$ to get some perspective about the quantum and thermal melting of the electronic/excitonic crystals recently discovered in bilayer TMDs moiré patterns.

Throughout the paper we stick to the convention $e = \hbar = 4\pi\epsilon_0 = 1$, with $e$ the charge of an electron, $\hbar$ the reduced Planck constant and $\epsilon_0$ the vacuum permittivity. The energy scale that characterizes the crystal melting corresponds to the inter-site interaction strength. In 2D layered materials, the interaction between electrons corresponds to the Coulomb potential modified by the atomically-thin geometry of the layered structure, which can be expressed in the Rytova-Keldysh form[33]

$$V_C(\mathbf{r}) = \frac{\pi}{2\epsilon r_0}\left[H_0\left(\frac{r}{r_0}\right) - Y_0\left(\frac{r}{r_0}\right)\right], \tag{1}$$

where $H_0$ and $Y_0$ are the Struve and 2nd-kind Bessel functions, respectively. $r_0$ is the 2D screening length (larger $r_0$ implies stronger screening effect of the layered material), and $\epsilon$ is the relative dielectric constant of the environment. Different monolayer TMDs have similar screening lengths $r_0 \approx 5/\epsilon$ nm [34], while $r_0 \approx 10/\epsilon$ nm can approximately describe Coulomb interactions in bilayer TMDs. In the limit $r_0 \to 0$, $V_C(\mathbf{r}) \to \frac{1}{\epsilon r}$. Meanwhile, the interaction between IXs in bilayer structures is given by the electric dipolar repulsion $V_D(\mathbf{r}) = 2V_{\text{intra}}(\mathbf{r}) - 2V_{\text{inter}}(\mathbf{r})$, where $V_{\text{intra}}$ ($-V_{\text{inter}}$) corresponds to the intralayer Coulomb repulsion between two electrons or two holes (interlayer Coulomb attraction between an electron and a hole). Here for simplicity, we have modeled IXs as point dipoles because of the small Bohr radius ($\approx 2$ nm [35]). Modified by the 2D geometry of the bilayer structure, the Fourier transforms of $V_{\text{intra}}(\mathbf{r})$ and $V_{\text{inter}}(\mathbf{r})$ have the following forms[36]

$$\begin{aligned}V_{\text{intra}}(\mathbf{k}) &= \frac{2\pi}{\epsilon k}\frac{1 + r_0 k(1 - e^{-kd})(1 + e^{-kd})}{[1 + r_0 k(1 - e^{-kd})][1 + r_0 k(1 + e^{-kd})]},\\ V_{\text{inter}}(\mathbf{k}) &= \frac{2\pi}{\epsilon k}\frac{e^{-kd}}{[1 + r_0 k(1 - e^{-kd})][1 + r_0 k(1 + e^{-kd})]}.\end{aligned} \tag{2}$$

Here $d$ is the interlayer distance. In the above equations, we have assumed that the screening lengths $r_0$ of the two monolayers are the same. Obviously $V_{\text{intra}}(\mathbf{r}) \to \frac{1}{\epsilon r}$ and $V_{\text{inter}}(\mathbf{r}) \to \frac{1}{\epsilon\sqrt{r^2+d^2}}$ when $r_0 \to 0$, in this limit $V_D(\mathbf{r}) \to \frac{2}{\epsilon}\left(\frac{1}{r} - \frac{1}{\sqrt{r^2+d^2}}\right)$ converges to the traditional form in 3D homogeneous space. For a finite $r_0$ value, the dipolar interaction can be expressed as

$$V_D(\mathbf{r}) = \frac{1}{\epsilon\pi}\int d\mathbf{k}\frac{(1 - e^{-kd})e^{i\mathbf{k}\cdot\mathbf{r}}}{k(1 + r_0 k(1 - e^{-kd}))}. \tag{3}$$

Under $T = 0$, we use two approaches to obtain $\langle\mathbf{r}^2\rangle$ and the traditional Lindemann ratio of the triangular crystal. First approach is a mean-field treatment on the interaction between crystal sites, which can give an intuitive picture for the localization of the particle. Near a lattice site located at $\mathbf{R}_0 = 0$, a particle feels other particles' repulsion as well as the background moiré potential (see the illustration in Fig. 1(a)), the mean-field trapping potential can be written as

$$V_{\text{total}}(\mathbf{r}) = \sum_{n\neq 0}\int d\mathbf{r}_n|\psi(\mathbf{R}_n + \mathbf{r}_n)|^2 V(\mathbf{R}_n + \mathbf{r}_n - \mathbf{r}) + V_{\text{moiré}}(\mathbf{r}). \tag{4}$$

Here $\psi(\mathbf{R}_n + \mathbf{r}_n) = \frac{1}{\sqrt{\pi}\sigma}\exp\left(-\frac{r_n^2}{2\sigma^2}\right)$ is the single-particle trial wave function of the $n$-th site, which is assumed to be in a gaussian form with $\sigma$ the localization length and $\mathbf{R}_n$ the equilibrium position of this site. $V(\mathbf{R}_n + \mathbf{r}_n - \mathbf{r})$ is the interaction between two particles located at $\mathbf{R}_n + \mathbf{r}_n$ and $\mathbf{r}$, respectively, and $V_{\text{moiré}}(\mathbf{r})$ is the periodic moiré potential. We expand both the moiré and interaction potentials up to the 2nd-order of $r$, that is, $V_{\text{moiré}}(\mathbf{r}) \approx \frac{\gamma}{2\lambda^2} r^2$ and $V(\mathbf{R}_n + \mathbf{r}_n - \mathbf{r}) \approx V(\mathbf{R}_n) + (\mathbf{r}_n - \mathbf{r}) \cdot \frac{\partial V(\mathbf{R}_n)}{\partial \mathbf{R}_n} + \frac{1}{2}\left((\mathbf{r}_n - \mathbf{r}) \cdot \frac{\partial}{\partial \mathbf{R}_n}\right)^2 V(\mathbf{R}_n)$. Here $\lambda \sim 10$ nm is the wavelength of the crystal or the nearest-neighbor separation, $\gamma$ corresponds to the confinement strength at the moiré potential minima which is estimated to be in the order of 0.1-1 eV [37-39]. The trapping potential then has a harmonic form

$$V_{\text{total}}(\mathbf{r}) \approx E_0 + \frac{1}{2}\left(\frac{\gamma}{\lambda^2} + \frac{1}{\pi\sigma^2}\sum_{n\neq 0}\int d\mathbf{r}_n \frac{\partial^2 V(\mathbf{R}_n + \mathbf{r}_n)}{\partial \mathbf{R}_n^2} e^{-\frac{r_n^2}{\sigma^2}}\right) r^2. \tag{5}$$

This results in

$$\langle \mathbf{r}^2 \rangle = \sigma^2 \approx m^{-\frac{1}{2}}\left(\frac{\gamma}{\lambda^2} + \frac{1}{\pi\sigma^2}\sum_{n\neq 0}\int d\mathbf{r}_n \frac{\partial^2 V(\mathbf{R}_n + \mathbf{r}_n)}{\partial \mathbf{R}_n^2} e^{-\frac{r_n^2}{\sigma^2}}\right)^{-\frac{1}{2}}. \tag{6}$$

In the above equation, $m$ is the effect mass of the particle, $E_0 \equiv \frac{1}{2}\sum_{n\neq 0} V(\mathbf{R}_n)$ corresponds to a constant potential. The localization length $\sigma$ can then be solved self-consistently.

Our second approach is to introduce the creation (annihilation) operator $\hat{a}^\dagger_{l,\mathbf{k}}$ ($\hat{a}_{l,\mathbf{k}}$) for $l$-th phonon branch with a wave vector $\mathbf{k}$, and write the system Hamiltonian as[40]

$$\begin{aligned}\hat{H} &= \sum_n \left(-\frac{1}{2m}\frac{\partial^2}{\partial \mathbf{r}_n^2} + V_{\text{moiré}}(\mathbf{R}_n + \mathbf{r}_n) + \frac{1}{2}\sum_{n'\neq n} V(\mathbf{R}_{n'} - \mathbf{R}_n + \mathbf{r}_{n'} - \mathbf{r}_n)\right) \\ &\approx \sum_{\mathbf{k}}\left(E_0 + \sum_l \omega_{l,\mathbf{k}}\left(\hat{a}^\dagger_{l,\mathbf{k}}\hat{a}_{l,\mathbf{k}} + \frac{1}{2}\right)\right).\end{aligned} \tag{7}$$

This gives a rigorous expression for the mean-square-displacement

$$\langle \mathbf{r}^2 \rangle = \frac{\Omega}{(2\pi)^2 m}\sum_l \int d\mathbf{k}\frac{\langle \hat{a}^\dagger_{l,\mathbf{k}}\hat{a}_{l,\mathbf{k}}\rangle + 1/2}{\omega_{l,\mathbf{k}}}. \tag{8}$$

Note that the phonon occupation $\langle \hat{a}^\dagger_{l,\mathbf{k}}\hat{a}_{l,\mathbf{k}}\rangle = 0$ in the zero-temperature limit, so the traditional Lindemann ratio $\eta \equiv \sqrt{\langle \mathbf{r}^2 \rangle}/\lambda$ can be obtained from the phonon dispersion.

In obtaining Eq. (6) and (8), we have used the harmonic approximation which requires the hopping between moiré potential minima to be suppressed by the interaction, that is, the strong correlation limit. On the other hand, a weak correlation limit can be realized in TMDs moiré patterns under a filling factor $v \ll 1$, where the distance between the crystal sites is much larger than that between the nearest-neighbor moiré potential minimum. In this case, the hopping from an occupied moiré potential minimum to an empty one cost very small interaction potential thus cannot be ignored. This implies that each electron/exciton site will cover multiple moiré potential minima, resulting in $\langle \mathbf{r}^2 \rangle$ significantly larger than those in Eq. (6) and (8). To calculate the correct $\langle \mathbf{r}^2 \rangle$ value in the weak correlation limit, one can first solve the single-particle mini-bands under the periodic moiré potential to obtain the effective mass $m_{\text{moiré}}$ at the energy minimum. Then the electrons/excitons are treated as particles with mass $m_{\text{moiré}}$ which weakly interact with each other, and Eq. (6) and (8) with $\gamma = 0$ can be used to get the Lindemann ratio. However, the inevitable disorder potential will dominate over the interaction under $v \ll 1$, which introduces fluctuations and can break the long-range crystalline order[41,42]. Below we focus on the strong correlation limit, and assume that the harmonic approximation is always valid.

For the electronic crystal with $V(\mathbf{r}) = V_C(\mathbf{r})$ (Eq. (1)), we show our calculated $\eta$ as a function of the moiré wavelength $\lambda$ in Fig. 1(b-d), for several values of environmental screening $\epsilon$, screening length $r_0$ and moiré trapping strength $\gamma$. Unless specified, we set the electron effective mass as $m = 0.5 m_0$ with $m_0$ the free electron mass. The results obtained from Eq. (6) and Eq. (8) are shown as solid and empty symbols, respectively, which show qualitative agreement. The rigorous results of $\eta$ from the phonon dispersion are generally larger than the mean-field results (especially when $\gamma = 0$), implying that the inter-site correlation not taken into account in the mean-field treatment can lead to delocalization[43]. Fig. 1(b) and (c) correspond to the case of a suspended TMDs layer ($\epsilon = 1$), whereas Fig. 1(d) simulates a structure encapsulated by thick hBN layers ($\epsilon = 5$). Note that $m = 0.8 m_0$ is close to the measured effective mass of **K**-valley electrons in monolayer MoSe$_2$[44]. Just as expected, larger $r_0$ values lead to stronger 2D screening thus weaker localization (or larger $\eta$). For a given $r_0$ value, $\eta$ decays with the increase of $\lambda$, implying that the crystal phase is favored under a low electron density. This can be understood from the mean-field result in Eq. (6): for large values of $\lambda$, $\eta$ scales as $\lambda^{-1/4}$ when $\gamma = 0$, but scales as $\lambda^{-1/2}$ for a finite $\gamma$. According to the Lindemann criterion, the electronic crystal will melt into a liquid when $\eta$ is above $\eta_c \approx 0.2\text{-}0.25$ [28,29]. From Fig. 1(b-d), we can see that a strong moiré confinement $\gamma$ can greatly facilitate the formation of the crystal phase with a short wavelength $\lambda \sim 10$ nm, especially when the environmental screening $\epsilon$ is large (Fig. 1(d)).

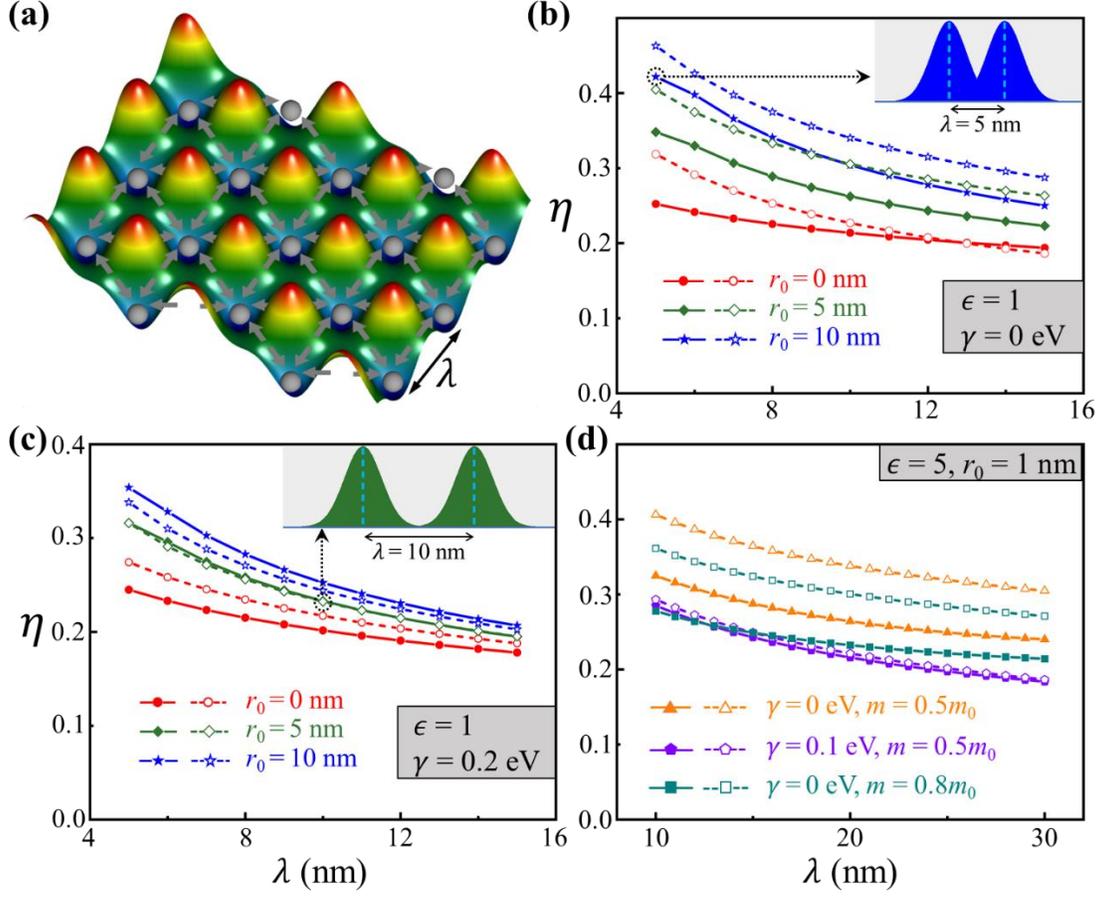

**Figure 1.** (a) Schematic illustration of an electronic triangular crystal in a moiré superlattice with wavelength $\lambda$. The colored landscape shows the moiré potential, and the electrons (gray balls) are trapped at the potential minima. The arrows between the balls denote their mutual Coulomb repulsions. (b-c) The calculated traditional Lindemann ratio $\eta$ under different sets of parameters, for an electron effective mass $m = 0.5m_0$ and environmental screening $\epsilon = 1$. The solid and empty symbols are the results obtained from the mean-field treatment (Eq. (6)) and the phonon dispersion (Eq. (8)), respectively. The insets show the overlap between two nearest-neighbor wavepackets under the mean-field treatment, for the chosen data points. (d) $\eta$ as a function of $\lambda$ under $\epsilon = 5$ and $r_0 = 1$ nm, which simulates TMDs encapsulated by thick hBN layers.

For the excitonic crystal, we show the calculated interaction strength $V(\mathbf{r}) = V_D(\mathbf{r})$ as a function of $\mathbf{r}$ in Fig. 2(a) under $d = 0.6$ nm and several values of $r_0$. Unlike the Coulomb interaction between electrons whose strength is always weakened by increasing $r_0$, the dipolar interaction between IXs shows a complicated behavior with $r_0$. As shown in Fig. 2(a), the strength of $V_D(\mathbf{r})$ gets enhanced (weakened) by the 2D screening when $r \sim r_0$ ($r \ll r_0$). As will be shown below, such an enhancement can facilitate the formation of the crystal phase. For $r \gg r_0$, the dipolar interaction converges to the traditional form $V_D(\mathbf{r}) \approx \frac{d^2}{r^3}$ in 3D homogeneous space.

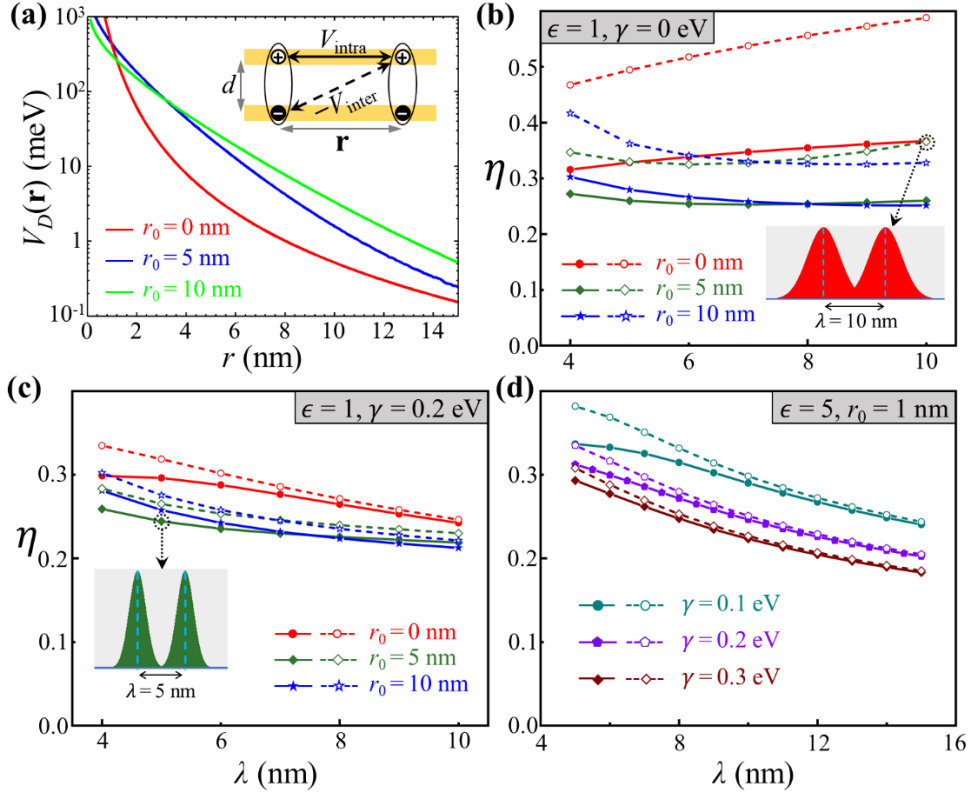

**Figure 2.** (a) The dipolar interaction strength between two IXs under $\epsilon = 1$ and $d = 0.6$ nm, for three different $r_0$ values. The inset illustrates the intralayer (interlayer) Coulomb interaction $V_{intra}$ ($-V_{inter}$) between the same charge (opposite charges). (b-d) The calculated Lindemann ratio as a function of $\lambda$, under different sets of parameters. The exciton effective mass is set as $m = m_0$. The solid and empty symbols are the results obtained from the mean-field treatment (Eq. (6)) and the phonon dispersion (Eq. (8)), respectively. The insets show the overlap between nearest-neighbor wavepackets for the chosen data points.

The calculated traditional Lindemann ratio for the excitonic crystal with an IX effective mass $m = m_0$ is shown in Fig. 2(b-d), where the solid (empty) symbols correspond to results obtained from Eq. (6) (Eq. (8)). For the case of $\gamma = 0$ and $r_0 = 0$ (red line in Fig. 2(b)), $\eta$ becomes larger when increasing $\lambda$, implying that the crystal phase is favored under a small $\lambda$. However the obtained value of $\eta$ is quite large (especially the rigorous results from Eq. (8) which are significantly larger than the mean-field results), implying that a rather high IX density is needed to realize the crystal phase. However, when taking into account the finite Bohr radius ($\approx 2$ nm), excitons will undergo a Mott transition to the electron-hole plasma above a critical density, thus preventing the crystal phase to form under this set of parameters. Meanwhile, a finite $r_0$ can facilitate the formation of the crystal phase, as indicated by the very different behaviors of the $r_0 = 0$ and $r_0 \neq 0$ curves in Fig. 2(b). For $\gamma = 0$ and $r_0 \neq 0$, $\eta$ shows an anomalous behavior of first decreasing then slowly increasing with $\lambda$, exhibiting a minimum at a certain $\lambda$ value. Such a behavior is due to the fact that the dipolar interaction strength gets enhanced by the 2D screening in the regime $\lambda \sim r_0$ (see Fig. 2(a)). The presence of a finite moiré confinement $\gamma$ can further facilitate the formation of the crystal phase. In Fig. 2(c) with $\gamma = 0.2$ eV, $\eta$ becomes smaller when increasing $\lambda$, similar to that in the electronic crystal. Now the crystal phase with $\eta < \eta_c$ can be realized when $\lambda$ is sufficiently large. Fig. 2(d) corresponds to a large environmental screening ($\epsilon = 5$, $r_0 = 1$ nm), in this case a large $\gamma$ is essential for realizing the crystal phase.

We now consider the thermal melting of the electronic/excitonic crystal. In the absence of the moiré confinement ($\gamma = 0$), the long-wavelength transverse phonon mode has a linear dispersion $\omega_{T,\mathbf{k}\to 0} = ck$ with $c$ the group velocity[40]. Under a finite temperature $T$, $\langle \hat{a}^\dagger_{T,\mathbf{k}} \hat{a}_{T,\mathbf{k}} \rangle_{\mathbf{k}\to 0} = \frac{k_B T}{ck}$ and the mean-square displacement obtained from Eq. (8) diverges as $\langle \mathbf{r}^2 \rangle \propto T \ln S$, where $S$ is the area of the 2D material. This implies that the traditional Lindemann criterion fails to describe the thermal melting of the 2D Wigner crystal under $\gamma = 0$. Instead, here we consider the modified Lindemann ratio $\eta^{(m)} \equiv \sqrt{\langle (\mathbf{r}_n - \mathbf{r}_{n'})^2 \rangle}/\lambda$ which focuses on the fluctuation of the relative displacement for the nearest-neighbor sites $(n, n')$. In the triangular crystal, we use the phonon operators to write

$$\langle (\mathbf{r}_n - \mathbf{r}_{n'})^2 \rangle = \frac{\hbar \Omega}{(2\pi)^2 m} \sum_l \int d\mathbf{k} \frac{2\langle \hat{a}^\dagger_{l,\mathbf{k}} \hat{a}_{l,\mathbf{k}} \rangle + 1}{\omega_{l,\mathbf{k}}} (1 - \cos(\mathbf{k} \cdot \mathbf{R}_{nn'})). \qquad (9)$$

Here $\mathbf{R}_{nn'} \equiv \mathbf{R}_n - \mathbf{R}_{n'}$ is the relative displacement between the nearest-neighbor sites, and $\langle \hat{a}^\dagger_{l,\mathbf{k}} \hat{a}_{l,\mathbf{k}} \rangle = \left( \exp\left(\frac{\omega_{l,\mathbf{k}}}{k_B T}\right) - 1 \right)^{-1}$. Note that $\langle (\mathbf{r}_n - \mathbf{r}_{n'})^2 \rangle$ is finite even for the $\gamma = 0$ case under $T > 0$. Below we use $\eta^{(m)}$ obtained from Eq. (9) to investigate the thermal melting of electronic and excitonic crystals.

Fig. 3(a) is the schematic illustration of a triangular electronic crystal, the corresponding $\eta^{(m)}$ values as functions of $\lambda$ and $T$ under different sets of system parameters are shown in Fig. 3(b). $\eta^{(m)}$ increases with the increasing of $T$ and the decreasing of $\lambda$, and depends sensitively on the moiré confinement strength $\gamma$. We note that an electronic Wigner crystal with a wavelength $\lambda \approx 27$ nm has been observed in hBN-encapsulated monolayer MoSe$_2$ below 11 K[10], which is in qualitative agreement with the $\gamma = 0$ result in Fig. 3(b). The quantitative disagreements in the range of $\lambda$ and $T$ for the crystal phase could be due to the discrepancy in $\eta_c^{(m)}$ and system parameters like $\epsilon$, $r_0$ and $m$. For finite $\gamma$ values, the electronic crystal can form at $\lambda \sim 10$ nm and below a critical temperature which ranges from several tens Kelvin to above 100 K (depending on parameters like $\gamma$, $\epsilon$, $r_0$ and $m$). This is also in qualitative agreement with the experiment in Ref.

[8], where electronic crystals under $v = 1$ in WSe2/WS2 moiré patterns have been observed below ~150 K. We emphasize that the critical value $\eta_c^{(m)}$ obtained using different method can be different, thus our result of $\eta^{(m)}$ can only give a qualitative estimation to the melting temperature. Meanwhile, $\eta_c^{(m)}$ is also temperature-dependent. For instance, in the classical limit such that the thermal fluctuation dominates over the quantum fluctuation (the high-temperature and low-density case), the critical value becomes $\eta_c^{(m)} \approx 0.15$ [32]. Due to the small effective mass of the electron or IX, here we focus on the limit that the quantum fluctuation dominates over the thermal fluctuation, and ignore the temperature-dependence of $\eta_c^{(m)}$.

Electronic crystals other than the triangular type can form under a fractional filling factor, including the linear-stripe ($v = 1/2$), zigzag-stripe ($v = 1/2$) and honeycomb ($v = 2/3$) crystals[4,8,9,11,13,15]. We have calculated $\eta^{(m)}$ for these electronic crystals using the harmonic approximation, where $\langle (\mathbf{r}_n - \mathbf{r}_{n'})^2 \rangle$ is obtained from an equation similar to Eq. (9) and the phonon dispersions have been calculated in Ref. [40]. Fig. 3(c,e,g) are the schematic illustrations of the linear-stripe, zigzag-stripe and honeycomb electronic crystals, respectively, and Fig. 3(d,f,h) show the corresponding $\eta^{(m)}$ values. Note that these types of electronic crystals are dynamically stable only under a large enough $\gamma$ (certain phonon modes have imaginary frequencies under weak $\gamma$ values) [40], thus only the results with finite $\gamma$ are shown. Compared to the triangular crystal, the lower densities of the linear-stripe, zigzag-stripe and honeycomb crystals decrease the Coulomb interaction strength and result in larger $\eta^{(m)}$ values. We note that in principle $\eta_c^{(m)}$ should vary with the lattice type. It is found that the honeycomb-type bilayer Wigner crystal observed in MoSe2/hBN/MoSe2 heterostructure exhibits a significantly larger modified Lindemann ratio compared to $\eta_c^{(m)}$ of the triangular crystal, where the crystal lattice constant is ≈ 6 nm and two subsites are located in opposite layers separated by 1.6 nm [45]. Nevertheless, we can get some perspective from how the calculated $\eta^{(m)}$ depends on $\lambda$, $T$ and other parameters. For the linear- and zigzag-stripe crystals formed under $v = 1/2$, the linear-stripe crystal has a significantly larger $\eta^{(m)}$ than the zigzag-stripe one under a weak $\gamma$ or small $\lambda$ (see Fig. 3(d,f)), implying that the latter can be realized more easily. On the other hand, all lattice types have similar $\eta^{(m)}$ values for a large $\gamma$ or large $\lambda$, because in this limit the carrier localization is dominated by the moiré confinement and the inter-site interaction plays a minor role. We note that in Ref. [8], the electronic crystals under fractional fillings are found to have melting temperatures significantly lower than that under $v = 1$ (~30 K vs. ~150 K). Meanwhile for different triangular crystals formed under $v = 1$, 1/3, 1/4 and 1/7, the corresponding melting temperatures are lower for smaller $v$. These could be due to the following reasons that are ignored when we use Eq. (9) to calculate $\eta^{(m)}$: (1) The disorder in the system can suppress the long-range crystalline order[41], whose effect is significant when the inter-site interaction is weak. (2) When $v$ decreases from 1 to $\ll 1$, the hopping between nearest-neighbor moiré potential minima becomes more and more important, which leads to larger $\eta^{(m)}$ values than our calculated results.

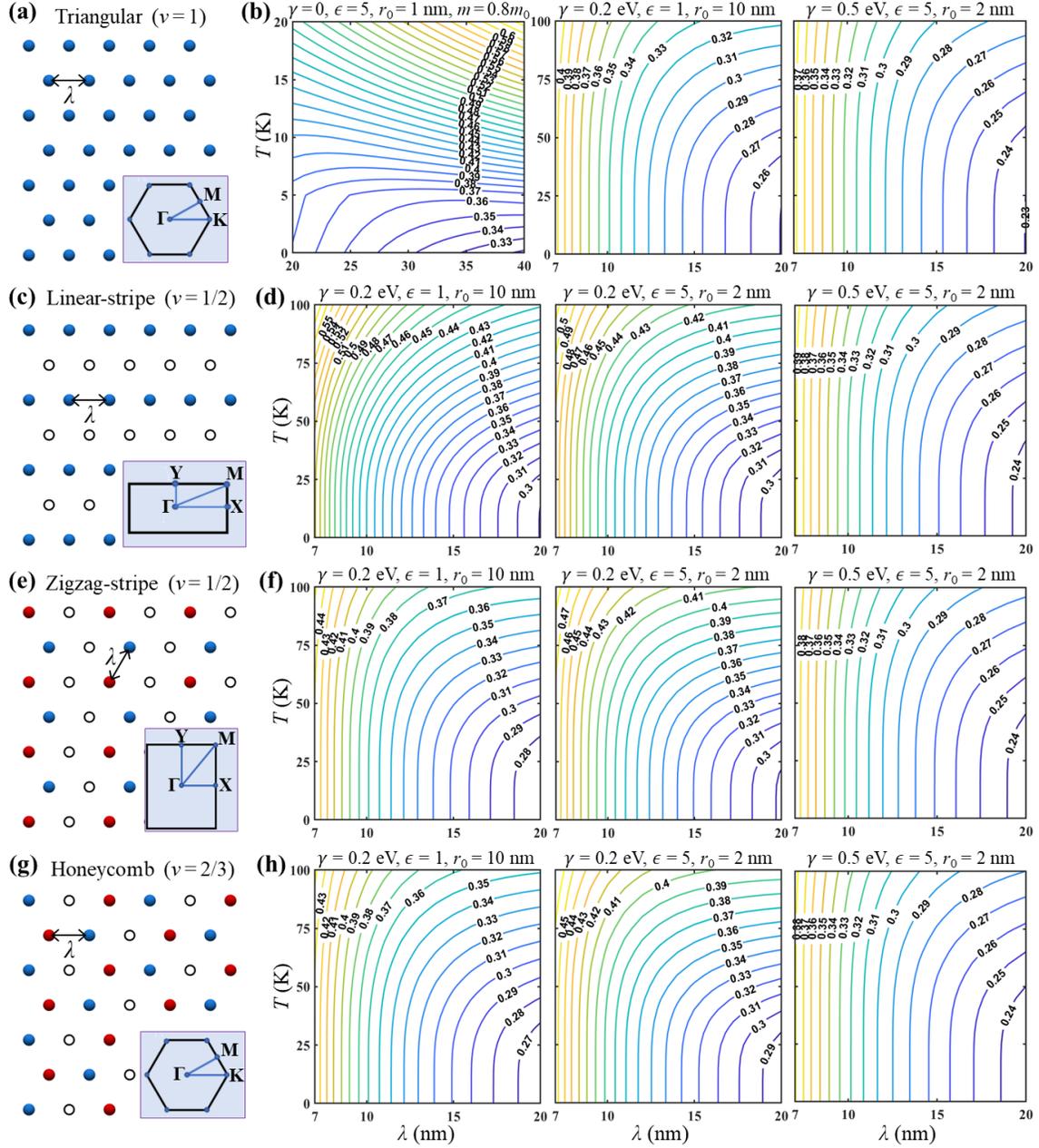

**Figure 3.** (a) Schematic illustration of the triangular electronic crystal under $v = 1$ and its Brillouin zone. (b) The contour plots of $\eta^{(m)}$ with $\lambda$ and $T$ for triangular electronic crystals, under various system parameters. (c) The linear-stripe electronic crystal under $v = 1/2$ and its Brillouin zone. (d) The contour plots of $\eta^{(m)}$ for linear-stripe crystals. (e) The zigzag-stripe electronic crystal under $v = 1/2$ and its Brillouin

zone. (f) The contour plots of $\eta^{(m)}$ for zigzag -stripe crystals. (g) The honeycomb electronic crystal under $v = 2/3$ and its Brillouin zone. (h) The contour plots of $\eta^{(m)}$ for honeycomb electronic crystals.

Fig. 4 is about the modified Lindemann ratio of excitonic crystals. Fig. 4(a,c,e,g) show our calculated phonon dispersions of the triangular, linear-stripe, zigzag-stripe and honeycomb excitonic crystals, respectively, under the parameters of $\epsilon = 1$, $\gamma = 0.2$ eV, $\lambda = 7$ nm, $d = 6$ Å. Due to the enhanced dipolar interaction strength, the variation range of the phonon frequency under $r_0 = 5$ or $10$ nm is significantly larger than that under $r_0 = 0$ nm. The corresponding $\eta^{(m)}$ values are shown in Fig. 4(b,d,f,h) as contour plots. Similar to the electronic crystal, when $\gamma = 0$ only the triangular-type excitonic crystal is dynamically stable with real phonon frequencies. In this case, $\eta^{(m)}$ has its minimum located at some finite $\lambda$. However, $\min(\eta^{(m)})$ is above $\eta_c^{(m)} \approx 0.31$ unless the dipolar interaction is quite strong (e.g., when the interlayer separation $d$ is large, see Fig. 4(b)). This is distinct from the electronic crystal where $\eta^{(m)}$ can always decrease to near zero for a large enough $\lambda$ (Fig. 3(b)). Note that in experiments, most of the observed IXs have $d \approx 0.6$ nm in bilayer TMDs or $d \approx 1$ nm in van der Waals stacked TMDs/hBN/TMDs system, where the dipolar interaction seems not strong enough to realize the excitonic crystal under $\gamma = 0$. On the other hand, under a finite moiré confinement ($\gamma > 0$) the behavior of $\eta^{(m)}$ for the excitonic crystal is qualitatively similar to that of the electronic crystal. In this case, the excitonic crystal can form under a large enough $\lambda$ and a low enough temperature. Again, we estimate that the melting occurs at a critical temperature ranging from several tens Kelvin at small $\gamma$ to above 100 K at large $\gamma$.

We note that in experiments, the excitonic crystals are observed only under integer fillings but not for fractional ones[24-27]. We expect that the formation of an excitonic crystal with a filling factor $v < 1$ will be much harder than the electronic crystal, because: (1) The dipolar interaction between excitons is much weaker than the Coulomb interaction between electrons, thus the effect of the disorder is more significant for the exciton case which can destroy the excitonic crystalline order. (2) In the $v \ll 1$ limit, IXs should be modeled as particles under $\gamma = 0$ but with an effective mass $m_{\text{moiré}}$ determined from the moiré mini-band, which weakly interact with each other through the dipolar potential. Fig. 3(b) (Fig. 4(b)) shows that $\eta^{(m)}$ of the electronic (excitonic) crystal decreases (increases) with the increase of $\lambda$ when $\gamma = 0$ when $\lambda$ is large. Thus in the $v \ll 1$ limit with vanishing disorder, the electronic crystal with a large $\lambda$ can always form, whereas the excitonic crystal cannot form.

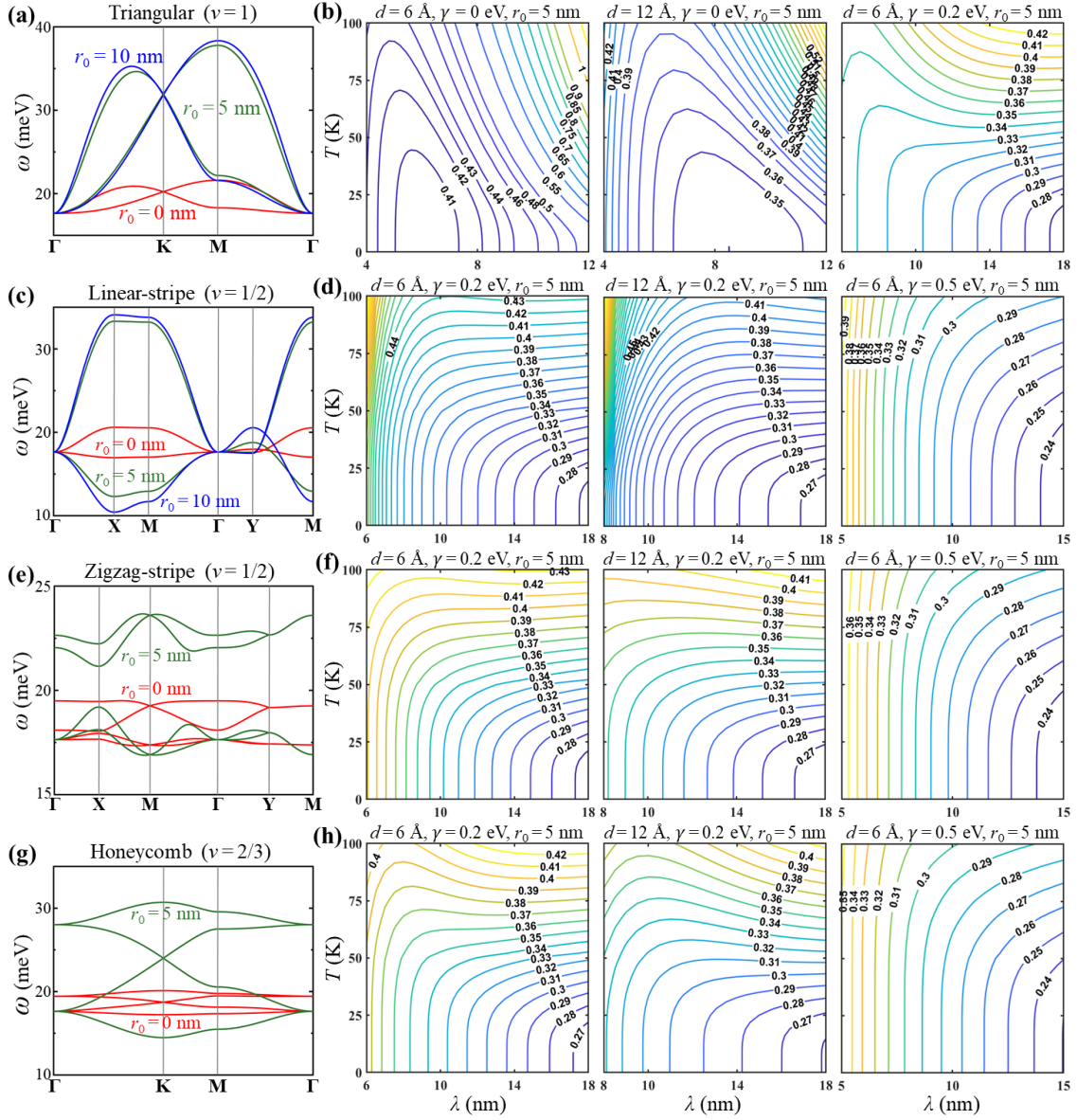

**Figure 4.** (a) The phonon dispersion of a triangular excitonic crystal under $\epsilon = 1$, $\gamma = 0.2$ eV, $\lambda = 7$ nm, $d = 6$ Å. (b) The contour plots of $\eta^{(m)}$ with $\lambda$ and $T$ for triangular excitonic crystals. (c) The phonon dispersion of a linear-stripe excitonic crystal under $\epsilon = 1$, $\gamma = 0.2$ eV, $\lambda = 7$ nm, $d = 6$ Å. (d) The contour plots of $\eta^{(m)}$ for linear-stripe excitonic crystals. (e) The phonon dispersion of a zigzag-stripe excitonic crystal under $\lambda = 7$ nm, $\epsilon = 1$, $\gamma = 0.2$ eV, $d = 6$ Å. (f) The contour plots of $\eta^{(m)}$ for zigzag-stripe excitonic crystals. (g) The phonon dispersion of a honeycomb excitonic crystal under $\epsilon = 1$, $\gamma = 0.2$ eV, $\lambda = 7$ nm, $d = 6$ Å. (h) The contour plots of $\eta^{(m)}$ for honeycomb excitonic crystals.

In summary, we have calculated the traditional and modified Lindemann ratios of electronic/excitonic crystals recently discovered in 2D semiconductor moiré patterns, which are used to analyze the quantum and thermal melting of these crystals. We find that in the absence of a moiré potential, the electronic crystal with a wavelength ~ 30 nm or larger can form under a temperature of tens Kelvin, whereas the excitonic crystal is very difficult to form in bilayer TMDs systems. A strong confinement from the moiré potential can facilitate the formation of electronic/excitonic crystals with a wavelength ~ 10 nm or smaller. Interestingly, the finite 2D screening from the atomically thin material can enhance the inter-site dipolar interaction, thus helps to realize the excitonic crystal. Our work can help to understand the requirement of the intriguing electronic/excitonic crystal phase in 2D semiconductor moiré patterns.

**Acknowledgements:** H.Y. acknowledges support by NSFC under grant No. 12274477, and the Department of Science and Technology of Guangdong Province in China (2019QN01X061).